\documentclass[12pt]{article}

\usepackage{amsmath,amssymb,graphicx,color} 
\usepackage{epsf}
\usepackage{pstricks}
\usepackage{cite}

\newcommand{\beq}{\begin{eqnarray}}
\newcommand{\eeq}{\end{eqnarray}}
\newcommand{\tabarr}[2]{\begin{array}{c} \textcolor{blue}{#1} \\ \textcolor{red}{#2} \end{array}}

\newcommand{\centeron}[2]{{\setbox0=\hbox{#1}\setbox1=\hbox{#2}\ifdim

\wd1>\wd0\kern.5\wd1\kern-.5\wd0\fi
\copy0

\kern-.5\wd0\kern-.5\wd1\copy1\ifdim\wd0>\wd1
                                       \kern.5\wd0\kern-.5\wd1\fi}}
\newcommand{\ltap}{\>\centeron{\raise.35ex\hbox{$<$}}
                               {\lower.65ex\hbox{$\sim$}}\>}
\newcommand{\gtap}{\>\centeron{\raise.35ex\hbox{$>$}}
                               {\lower.65ex\hbox{$\sim$}}\>}

\newcommand{\sech}{\mbox{sech}}

\newcommand\ZZ{\hbox{\zfont Z\kern-.4emZ}}
\font\zfont = cmss10 

\begin{document}
\begin{titlepage}
\begin{flushright}
{\tt hep-ph/0510366}
\end{flushright}

\vskip.5cm
\begin{center}
{\huge \bf Fully Radiative Electroweak Symmetry Breaking}

\vskip.1cm
\end{center}
\vskip0.2cm

\begin{center}
{\bf
{Giacomo Cacciapaglia}, {Csaba Cs\'aki}, and
{Seong Chan Park}}
\end{center}
\vskip 6pt

\begin{center}
{\it Institute for High Energy Phenomenology\\
Newman Laboratory of Elementary Particle Physics\\
Cornell University, Ithaca, NY 14853, USA } \\
\vspace*{0.3cm}
{\tt  cacciapa@mail.lns.cornell.edu, csaki@lepp.cornell.edu,
spark@lepp.cornell.edu}
\end{center}

\vglue 0.3truecm

\begin{abstract}
\vskip 3pt
\noindent

Models of Gauge-Higgs unification in extra dimensions offer a very elegant playground where one can study electroweak symmetry breaking.
The nicest feature is that gauge symmetry itself protects the Higgs potential from divergences, thus potentially curing the hierarchy problem in the Standard Model. In a flat space model tree-level contributions
to electroweak precision observables will be absent due to the flatness of the gauge and Higgs wave functions.
The Higgs
potential is fully radiatively generated and the contribution of bulk fermions will induce a vacuum expectation value.
A generic problem is that the quartic scalar coupling is too low, resulting in a Higgs VEV that is too close to the
compactification scale, and a Higgs mass that is too light.
In this paper we show that it is possible to solve these problems in a minimal scenario by cancellations in the Higgs potential between the contribution of different bulk fermions.
A crucial role is played by antiperiodic fermions: the cancellation is not the result of
a fine tuning, but rather dictated by the choice of representations and parities of the fermions.
We also show that introducing a relatively large representation can help in achieving a
sufficiently heavy top.
In this case, the strong coupling scale is lowered to a marginally acceptable value, and a more careful analysis of two loop effects should decide if the theory remains under perturbative control.

\end{abstract}

\end{titlepage}

\newpage


\section{Introduction}
\label{sec:intro}
\setcounter{equation}{0}
\setcounter{footnote}{0}

The Standard Model (SM) of electroweak interactions has withstood every recent experimental attempt
at  directly or indirectly detecting new physics.
On the other hand, it is widely believed that it is not the final theory but rather an effective theory valid up to a cutoff scale. Ultimately one would like to be able to find a theory where the cutoff scale is 
given by the Planck mass $M_{Pl} \sim 10^{19}$ GeV. The reason why the cutoff in the SM can not be very high is 
that loop corrections to the Higgs mass are quadratically sensitive to the cutoff of the theory. Thus new physics is expected to enter at a scale not too far from the weak scale to shield the Higgs mass 
from  this sensitivity to the physics at the Planck scale. This is usually referred to as the (big) hierarchy problem.
A qualitatively different tension arises when one takes into account the precision measurements that suggest a light Higgs boson, below $280$ GeV:
due to the hierarchy problem discussed above this would imply a low scale for the new physics around $1$ TeV.
This limit comes mainly from the top loop contributions to the Higgs mass, as the top Yukawa is the largest coupling to the Higgs. On the other hand, if one includes higher dimensional operators (which would presumably be generated
in a theory beyond the SM) in the analysis
of electroweak precisions observables~\cite{littlehierarchy}, it turns out that the scale suppressing such operators, that is the new physics scale, has to be larger than $5-10$ TeV.
This factor of $5-10$ mismatch is the little hierarchy problem: it requires a few \% fine tuning in the new physics contribution to the Higgs mass.
Thus, the precision measurements, mainly performed at LEP~\cite{LEP}, are the main challenge for the realization of realistic models of New Physics beyond the SM~\cite{LEP2}.

The most acknowledged paradigm addressing these problems is supersymmetry: due to the presence of partners of any SM particle with different spin, the dangerous quadratic dependence on the cutoff is tamed to a mild logarithmic dependence.
However, a natural realization of this paradigm would require superpartners at the weak scale, more precisely around the mass of the weak gauge bosons.
The fact that superpartners have not been discovered at LEP and the Tevatron 
has pushed up the viable scale of supersymmetry breaking.
In a minimal realization (the MSSM), this reintroduces the fine tuning problem, especially in the stop mass term
which is  responsible both for EWSB and the taming of the top divergences.

The Large Hadron Collider (LHC), which is expected to turn on in just two years, is expected  to shed some
experimental light on the problem. This adds a great motivation to look for alternative models that naturally
protect the Higgs mass.
Indeed a plethora of new mechanisms have been proposed, among them gauge extensions of the MSSM~\cite{gaugeext},
Little Higgses~\cite{littlehiggs}, fat Higgses~\cite{fathiggs}, and many more.
The recent realization~\cite{antoniadis} that extra dimensions could play a role in low energy physics and are
not necessarily only relevant at the Planck scale, has opened a Pandora box of new possibilities.
An incomplete list of such models (relevant to electroweak symmetry breaking)
includes the Randall-Sundrum model~\cite{RS},
extra dimensional supersymmetry~\cite{BHN}, composite Higgs in warped space~\cite{contino1,contino2}, and Higgsless models~\cite{higgsless}.

A very attractive idea utilizing extra dimensions is called
Gauge-Higgs unification, and was first discussed in~\cite{gaugehiggs}, and then
developed by several authors both in 5~\cite{5D,SSS,twisted} and 6
dimensions~\cite{ABQ,CGM,6D-1,tadpoles}. In a nutshell, the idea is
to identify the Higgs as the component along the extra dimensions of
a gauge field. The residual gauge invariance after the orbifold
breaking will impose a shift symmetry on the Higgs. The potential
yielding electroweak symmetry breaking is then radiatively generated
and gauge invariance itself, embedded in the extra dimensional
background, ensures the insensitivity of the Higgs mass and the
electroweak scale on the cutoff. In 6D, however, possible tadpoles
can be generated on the orbifold fixed points~\cite{CGM,tadpoles},
reintroducing the cutoff dependence on the Higgs mass. While this
mechanism offers great simplicity and elegance, any effort to build
a realistic model up to date has been unfruitful. The main problems
are the lightness of the Higgs and of the top quark. Regarding the
top, the Yukawas are generated via the gauge coupling itself, so it
is generically hard to engineer a Yukawa of order 1 from a small
gauge coupling. Regarding the Higgs mass, it turns out to be too
small, below the value currently excluded by LEP, because the
quartic scalar interaction term is generated at one loop. Since the
entire potential (mass and quartic) is loop generated, the potential
will also generically prefer large values of the Higgs vacuum expectation value (VEV) relative
to the compactification scale so that the scale of new physics stays
dangerously low.
It is interesting to note that a deconstructed version of this mechanism~\cite{dec} led to the idea of Little Higgs models.
The symmetry protecting the Higgs mass is now a discrete shift symmetry, and the construction is much less constrained by the absence of 5D Lorentz invariance.
In Little Higgs models, this idea has been pushed further: in this case the symmetry is protecting the Higgs mass at one loop, but allows a quartic coupling at tree level~\cite{littlehiggs}.

The simplest possibility is to extend the weak gauge symmetry to SU(3)$_w$~\cite{ABQ}, the smallest group
that allows to embed the Higgs in an adjoint representation together
with the SM gauge bosons. The SU(3)$_w$ is broken to the SM gauge
group SU(2)$_L \times$U(1)$_Y$ by an orbifold projection. Such model, in one flat extra dimension,
has been extensively studied in~\cite{SSS}: in their numerical study
the authors indeed found typical values $m_H < 20 $ GeV and $m_t <
100$ GeV. An interesting possibility to improve the situation is to
warp the extra dimension~\cite{contino2}: this allows to raise both
the Higgs and top mass. However, in such background, the Higgs VEV
distorts the W and Z wave functions and generates corrections to the
couplings with fermions. Thus, one has to worry about tree level
oblique corrections, in particular protecting the $\rho$ parameter
requires the inclusion of a custodial SU(2)$_R$ symmetry in the
bulk~\cite{custodial}. Another crucial parameter is the coupling of
the left-handed bottom to the Z: forcing a large top mass will
induce large corrections, due to the fact that the b$_l$ is part of
the same weak doublet as the top~\cite{Zbb}. 
From this point of view, the flat case offers a great advantage.
Indeed, the flatness of the Higgs VEV does not induce any tree level mixing between the KK modes, due to the orthogonality of the wave functions.
The zero modes will also generically have flat wave functions, thus forbidding couplings with one KK mode alone: this property is similar to the T-parity introduced in Little Higgs models~\cite{Tparity}.

In this paper we will focus on a flat model, in particular the toy model analyzed in~\cite{SSS}, where such tree level corrections are avoided, thanks to the flatness of the $W$ and $Z$ wave functions.
We show two mechanisms that allow to get a heavy top and a heavy Higgs mass in a minimal scenario.
The fermion masses are generated by the mixing of localized degrees of freedom that couple to a massive bulk
field~\cite{CGM}.
A minimal model would require at least one bulk field for each SM fermion.
We have re-analyzed the spectrum of the top quark tower, taking into account the
effects of the localized mixings exactly.
Our numerical analysis confirms the results in~\cite{SSS}: the effective Yukawas are exponentially suppressed by the
bulk masses of the bulk fermions or by small localized mixings.
However, in the limit of vanishing bulk mass the light mode develops a mass of the order of the $W$ mass.
This limit corresponds to the case when the SM top is mostly a bulk degree of freedom.
An enhancement factor can be added if one considers larger representations: as a drawback adding large representations will also lower the scale where the theory enters a strong coupling regime, endangering the stability of the calculation.
We find that a relatively small representation may explain a factor of two between the top and the $W$ and still keep the theory under control. 
However, a more careful analysis of this issue is necessary.
In order to get rid of unwanted extra light modes, one can either add localized degrees of freedom on the brane and couple them with the zero modes via localized mass terms, or twist the fermionic boundary conditions and then introduce the SM fermions on the brane.

The second problem is the Higgs mass.
In~\cite{SSS}, the Higgs is found to be too light, however only the contribution of the top quark tower is taken into account in calculating the Higgs potential.
The rationale behind this is the following: since all the other SM fermions are much lighter than the top,
one expects the corresponding bulk fermions to be heavier in order to naturally generate the hierarchies in the
fermion spectrum.
However, if one gives up this motivation and simply assumes that the hierarchies are generated by smaller mixings,
other bulk fermions, that couple to the bottom, tau or light generations, can give a non-negligible contribution to
the Higgs potential.
We identify a set of scenarios where these contributions show cancellations in the potential, achieving two main goals: first of all the Higgs VEV in units of 1/$R$ is much smaller, thus allowing the resonance scale to be much heavier than the W mass, and the Higgs mass receives additional contributions that push its value above the experimental bound.
It is important to notice that these cancellations are not to be considered a fine tuning.
Indeed, they are generated by the contribution of different representations of the SU(3)$_w$ group and will not be spoiled by a continuous variation of the parameters involved.
Moreover, it is important for us that the model is still minimal, in the sense that we do not add any bulk ``spectator'' field with the sole purpose of raising the Higgs mass: each fermion we take into account has to be introduced in the theory anyway in order to generate the masses of light fermions.
An important ingredient for the model to work is the introduction of antiperiodic bulk fermions (twisted boundary conditions): the twisting flips the sign of the fermion contribution to the Higgs mass~\cite{twisted}, that in 4D is always negative, and this flipped sign is the main source for the cancellations in the potential.
However, in a fully realistic theory there will still be contributions to the $\rho$ parameter and to the coupling of the $Z$ boson with the bottom quark.
$\Delta \rho$ is due to the fact that we need to introduce an extra U(1) to obtain the correct $\sin^2 \theta_w$ and to get the correct quark hypercharges.
Once the extra U(1) is broken by boundary terms, the $Z$ will become a mixture of the $A^3$ and $A^8$ fields from SU(3)$_w$ and the extra U(1) field, with a non-flat wave function.
This distortion also generates a correction to $Z b \bar b$.
Another correction to the couplings of the bottom arises due to the presence of triplets in the representation containing the left-handed top (and bottom).
The orbifold projection will leave zero modes for those triplets, which have to be removed via localized mixing terms.
These triplet zero modes will couple to the quark doublet via the Higgs VEV.
In the end, a fine tuning comparable to that of the MSSM (few \% level) will still be required.

The paper is organized as follows: after a brief introduction of the
toy model in Sec.~\ref{sec:model}, we exhaustively discuss all the
possible ways of generating fermion masses in
Sec.~\ref{sec:fermmass}. In Sec.~\ref{sec:5Dmass} we illustrate how
one can avoid a light Higgs and large VEV in the toy model
of~\cite{SSS}. Finally, in Sec.~\ref{sec:topmass} we analyze how to
generate a realistic top mass using a large representation, and how
the same mechanism in the previous section can raise the Higgs mass. 
Finally we briefly discuss the bounds from precision measurements in Sec.~\ref{sec:EWPT}, before the conclusions and outlook in Sec.~\ref{sec:concl}.

\section{A toy SU(3)$_w$ model}
\label{sec:model}
\setcounter{equation}{0}
\setcounter{footnote}{0}

In this section we will briefly review the toy model studied in~\cite{SSS}.
The gauge group is SU(3)$_c \times$SU(3)$_w$ on an $S^1/Z_2$ orbifold: the enhanced weak symmetry allows the
unification of the SM gauge bosons and the Higgs doublet.
In fact, the adjoint of SU(3) decomposes into ({\bf 3}, 0) + ({\bf 2}, 1/2) + ({\bf 2}, -1/2) + ({\bf 1}, 0).
The orbifold breaks SU(3)$_w$ to SU(2)$_L \times$U(1)$_w$ via the projection matrix:

\begin{equation}
P = \left( \begin{array}{ccc}
-1 & 0 & 0\\
0 & -1 & 0\\
0 & 0 & 1
\end{array} \right)\,,
\end{equation}
where the gauge fields transform as $A_\mu \rightarrow P A_\mu P^\dagger$ and $A_5 \rightarrow - P A_5 P^\dagger$.
With this choice, only the SM gauge fields have a zero mode.
In the scalar sector, the zero mode is a single complex SU(2)
doublet with the correct quantum numbers to play the role of a Higgs:

\begin{equation}
A_5 = \frac{1}{\sqrt{2}} \left( \begin{array}{cc}
- &  H_5 \\
 H_5^\dagger & -
\end{array} \right)\,.
\end{equation}
Note also that all the massive modes in $A_5$ are eaten by the massive KK modes of the gauge bosons, and play the role of the longitudinal degrees of freedom, like in the usual Higgs mechanism: the only physical scalar left in the spectrum is the zero mode.
The linearized gauge transformations in the bulk are:

\begin{equation}
\begin{array}{lcr}
A_\mu & \rightarrow & A_\mu + \partial_\mu \lambda (x, x_5) + i [ \lambda (x,x_5), A_\mu ]\,, \\
A_5 & \rightarrow & A_5 + \partial_5 \lambda (x, x_5) + i [ \lambda (x,x_5), A_5 ]\,.
\end{array} \end{equation}
On the branes, $\lambda = 0$ for the broken generators, however the gauge transformation will still impose on $A_5$ a shift coming from $\partial_5 \lambda$.
This is enough to forbid a tree level potential for $A_5$, also on the fixed points, and only loop contributions will generate a potential for the Higgs, that will be non-local from the 5D point of view, and finite.
We will assume for the moment that the potential does induce a VEV for the Higgs: we can use SU(2) transformations to align the VEV, analogously to the SM case, and parametrize it

\begin{equation}
\langle H_5 \rangle = \sqrt{2}  \left( \begin{array}{c}
0\\
\alpha/R \end{array} \right)\,.
\end{equation}
It is now straightforward to compute the spectrum of the gauge bosons: we find

\begin{equation} \label{spectrum}
M_{W n} = \frac{n+\alpha}{R}\,, \quad M_{Z n} = \frac{n + 2 \alpha}{R}\,, \quad M_{\gamma n} = \frac{n}{R}\,,
\end{equation}
where $n \in Z$, and we want to identify the lightest state in each tower with the SM gauge bosons, the photon, the $W$ and the $Z$.
Let us first point out that the spectrum is invariant if we shift $\alpha$ by an integer, and if we change its sign.
In other words, the physical range for $\alpha$ is $[0,1/2]$ and all other vacua outside this range are equivalent, as the radiatively induced potential will respect the same symmetries.
Another important feature is that $M_Z$ turns out to be twice the $W$ mass: this is a consequence of the gauge group SU(3) that predicts $\theta_W = \pi/3$.
One possible way to fix it is to add localized gauge kinetic terms: SU(3) being broken on the boundaries, such terms can be different for the SU(2) and U(1) and, if large enough, can dominate and fix the correct value of $\sin \theta_W$.
However, this scenario is equivalent to a warped extra dimension: integrating out a slice of the warped space near the Planck brane, where the warping is small, will mimic the localized kinetic terms, while the remaining space will be almost flat.
We will not pursue this
direction, as it has been already discussed in the literature~\cite{contino2}. 
Moreover, it suffers from tree level corrections to the precision observables~\cite{Zbb} from the mixing of the zero modes with the KK modes as a consequence of the non-flat profile of the Higgs.
Another possibility is to extend the gauge group with an extra U(1)$_X$.
In this case, if the bulk fermions are charged, only the combination of the two U(1)'s proportional to the hypercharge is anomaly free, and the orthogonal gauge boson will develop a mass~\cite{ABQ}.
Alternatively, one can use boundary conditions to break U(1)$_w \times$U(1)$_X \to$ U(1)$_Y$,
for instance by twisting the BC on one of the two branes, such that no zero mode is left in the scalar sector.
In both cases, the breaking is due to localized terms: as a result the wave function of the $Z$ is distorted, introducing corrections to the $\rho$ parameter and $Z b \bar b$.
Finally, it might be possible to achieve the correct weak mixing angle starting from a different gauge group and using more complicated orbifold projections, thus without introducing distortions in the zero mode wave functions: this possibility has not been exaustively explored yet.
Nevertheless, the details of this mechanism will not affect the main results of this paper, so in the following we will assume the presence of the extra U(1).

\subsection{Bulk fermions}

The next problem is how to generate a mass for the SM matter fields.
If we added bulk fermions, with chiral zero modes thanks to the orbifold projection, the Higgs VEV would generate a
spectrum similar to that in (\ref{spectrum}): all the light modes would have masses larger than the $W$ mass, where the exact relation depends on group theory factors arising from the fermion representations.
Indeed, gauge invariance forces the Higgs to couple to bulk fields and with strength determined by the 5D gauge coupling $g_5$.
There are two possible solutions: one is to include odd masses for these fermions, that will localize the zero modes towards the two fixed points.
As modes with different chirality will be localized towards different points, this mechanism will reduce the overlaps between the wave functions, and generate hierarchies between the various Yukawa couplings.
Another possibility, adopted in~\cite{CGM,SSS} is to localize the SM fermions on the fixed points, and then mix them with massive bulk fields that will induce an effective Yukawa coupling {\it a la} Froggatt-Nielsen.
In the following we will focus on the latter possibility.

The most general conditions on a bulk fermion in a representation $\mathcal{R}$ are:

\begin{equation}
\psi (-y) = \eta \mathcal{R} (P) \psi (y)\,, \qquad \psi (2 \pi R+y) = \eta' \psi (y)\,.
\end{equation}
The presence of an extra parity $\eta'$ only means that we will allow for antiperiodic fermions: let us first discuss the case of periodic fermions, the case considered in detail in~\cite{SSS}.
In this case, the orbifold projection will leave chiral bulk zero modes: in order to get rid of them, for each fermion $\Psi$, we add a second bulk fermion $\tilde \Psi$ with the same quantum numbers but opposite parity, so that we can write down an invariant bulk mass $M$ for them.
Now, the localized fermions can mix with the even components that do not have vanishing wave functions on the fixed points.
In Table~\ref{tab:bulkf}, we listed the bulk fields, with their parities $\eta$ and SU(3)$_c \times$SU(2)$_L \times$U(1)$_Y$ decomposition, that contain components with the same quantum numbers as the SM fields.
Of course, in the presence of an extra U(1)$_X$, the extra charge can be adjusted to fit the hypercharge, and the choice of representations is much less constrained.

\begin{table}[t]
\centering
\begin{tabular}{|c|c|c|l|}
\hline
SM particle & SU(3)$_c \times$SU(3)$_w$ & $\eta$ & SU(3)$_c \times$SU(2)$_L \times$U(1)$_Y$ \\
\hline
down & ({\bf 3},{\bf 3}) & + & ({\bf 3},{\bf 1},-1/3) + ({\bf 3},{\bf 2},1/6) \\
up & ({\bf 3}, $\bf \bar 6$) & + & ({\bf 3},{\bf 1},2/3) + ({\bf 3},{\bf 2},1/6) + ({\bf 3},{\bf 3},-1/3)  \\
lepton & ({\bf 1},{\bf 10}) & + & ({\bf 1},{\bf 1},-1) + ({\bf 1},{\bf 2},-1/2) + ({\bf 1},{\bf 3},0) + ({\bf 1},{\bf 4},1/2)  \\
\hline
\end{tabular}
\caption{List of the bulk fermions that mix with the localized SM fermions.}
\label{tab:bulkf}
\end{table}

Antiperiodic fermions are equivalent to fermions with different parities on the two fixed points, that we will call ``twisted'', there is no massless zero mode and the KK masses are given by $m_n = (n+1/2)/R$.
In general, we can also add a partner $\tilde \Psi$ and a bulk mass, as in the previous case: such term could be of phenomenological interest, as we will see later.
Such twisted fermions can also account for the SM field masses.
Indeed, there will be components that do not vanish on the fixed points and can mix with localized fields.

\subsection{Higgs potential from the bulk fields}

We now discuss the contribution to the Higgs potential from these bulk fields.
Their spectrum, as a function of the Higgs VEV, generically takes the form:

\begin{equation}
m_n^2 = \frac{(n + \beta)^2}{R^2}\,, \quad n  \in Z\,,
\end{equation}
where the parameter $\beta$ is proportional to the Higgs VEV $\alpha$ via an integer, that is determined by the representation of the field.
We can use the Higgs-dependent spectrum to compute the full one-loop potential, using the Coleman-Weinberg formula: after summing over the KK modes~\cite{ABQ}, we find

\begin{equation} \label{eq:potgen}
V_{eff} (\beta) =  \frac{\mp 1}{32 \pi^2} \frac{1}{(\pi R)^4} \mathcal{F} (\beta)\,,
\end{equation}
where the signs stand for bosons/fermions and

\begin{equation}
\mathcal{F} (\beta) = \frac{3}{2} \sum_{n = 1}^{\infty} \frac{\cos (2 \pi \beta n)}{n^5} = \frac{3}{2} \mathcal{R}e \left[ \mbox{Li}_5 (e^{2\pi \beta i}) \right]\,,
\end{equation}
where Li$_k$ is the polylogarithmic function of order $k$.

In the presence of a bulk mass, as for example for the fermions described above, the spectrum is shifted to

\begin{equation}
m_n^2 = M^2 + \frac{(n + \beta)^2}{R^2}\,, \quad n \in Z\,,
\end{equation}
and the effective potential becomes

\begin{multline}
\mathcal{F}_\kappa (\beta) = \frac{3}{2} \sum_{n=1}^{\infty} \frac{e^{- \kappa n} \cos (2 \pi \beta n)}{n^{3}} \left( \frac{\kappa^2}{3} + \frac{\kappa}{n} + \frac{1}{n^2} \right) = \\
\frac{3}{2} \mathcal{R}e \left[ \frac{\kappa^2}{3} \mbox{Li}_3 (e^{-\kappa+2 \pi \beta i }) + \kappa \mbox{Li}_4
(e^{-\kappa+2 \pi \beta i }) + \mbox{Li}_5 (e^{-\kappa+2 \pi \beta i}) \right]\,,
\end{multline}
where $\kappa = 2 \pi M R$.
In the limit of vanishing bulk mass $\kappa \rightarrow 0$ we obtain the previous result, on the other hand for large $\kappa$ the contribution to the effective potential is exponentially suppressed.
As a consequence, the bulk fields with large bulk mass will contribute the less to the potential.
Another important feature is that the most important term in the series is a $\cos 2 \pi \beta$: the minimum of such term is in $0$ for bosons and in $\beta = 1/2$ for fermions.
Thus, the value of the integers relating $\beta$ to $\alpha$ will roughly speaking fix the value of the minimum.

If the fermion is antiperiodic or ``twisted'', the spectrum is:

\begin{equation}
m_n^2 = M^2 + \frac{(n + 1/2 + \beta)^2}{R^2}\,, \quad n  \in Z\,.
\end{equation}
The contribution to the effective potential is given by the previous formulas, with $\beta \rightarrow \beta+1/2$.
As
$$\cos (2 \pi n (\beta+1/2)) = (-1)^n \cos (2 \pi n \beta)\,,$$
the twisted parity approximately flips the overall sign of the contribution.
In this way, we can get positive contributions to the Higgs mass arising from fermions.

\begin{table}[t]
\centering
\begin{tabular}{|c|c|l|}
\hline
bulk field & multiplicity &  \\
\hline
gauge (adj.) & $- 3$ & $2 \mathcal{F} (\alpha) + \mathcal{F} (2 \alpha)$ \\
down (3) & $3\times 8$ & $\mathcal{F}_{\kappa_d} (\alpha)$ \\
up (6) & $3\times 8$ & $\mathcal{F}_{\kappa_u} (\alpha) + \mathcal{F}_{\kappa_u} (2 \alpha)$ \\
lepton (10) & $8$ & $2 \mathcal{F}_{\kappa_l} (\alpha) + \mathcal{F}_{\kappa_l} (2 \alpha)+ \mathcal{F}_{\kappa_l} (3 \alpha)$\\
\hline
\end{tabular}
\caption{Contribution to the Higgs potential from the bulk fields in the theory. The multiplicity counts the spin and color factors. Recall that for each SM fermion, there are 2 4-component bulk fermions.}
\label{tab:pot}
\end{table}

The contribution of each bulk field is now easily computed if we decompose the SU(3)$_w$ multiplets and compute the couplings to the Higgs.
The results are summarized in Table~\ref{tab:pot}.

\section{Fermion masses}
\label{sec:fermmass}
\setcounter{equation}{0}
\setcounter{footnote}{0}

In this section we will discuss how the SM fermions develop a mass
via the localized mixings. The bulk fields in Table~\ref{tab:bulkf}
contain a component with the same quantum numbers as the SM fields,
which thus they can mix with at the fixed points. The localized
Lagrangians generically are

\begin{multline} \label{eq:lloc}
\mathcal{L}_{loc} = \left[ -i {\bar Q}_L {\bar \sigma}^\mu \partial_\mu Q_L + \frac{\epsilon_L}{\sqrt{\pi R}} \psi^d Q_L + h.c. \right] \delta (y - y_L) +\\
\left[ -i q_R \sigma^\mu \partial_\mu {\bar q}_R + \frac{\epsilon_R}{\sqrt{\pi R}} q_R \chi^s  + h.c. \right] \delta (y - y_R)\,,
\end{multline}
where $\psi^d$ and $\chi^s$ are the doublet and singlet components of the bulk fermion and the mixing parameters $\epsilon$ are dimensionless (the factor of $\pi R$ has been chosen for future convenience).
A similar Lagrangian needs to be added for all the SM quarks and leptons.
The two points $y_L$ and $y_R$ can be either one of the fixed points: for each choice of twisted or untwisted periodicity, there is a unique component on both branes that can be identified with $\psi^d$ and $\chi^s$.
At the end of the day, we will have four inequivalent possibilities, depending on the twisting and if the mixings are on the same brane or on different brane.
In the following we will discuss all the cases, pointing out some differences.

In order to find how the spectrum is affected by the localized terms, we can as usual solve the bulk equations of motion in the bulk, and convert the localized terms into boundary conditions on the bulk fields~\cite{CGHST}.
Following this procedure, it is possible to write a master equation whose zeros are the mass eigenstates.
This procedure takes the effect of the localized mixings into account exactly.
For small mixings, the light mode mass is proportional to the $\epsilon$'s, so it is always possible to reproduce masses much smaller than the Higgs VEV.
The challenge is given by the top mass: the coupling of the Higgs is generated via gauge interactions, so it is not easy to achieve a Yukawa of order one from a much smaller gauge coupling.

In the following we will discuss the case of a fundamental for simplicity: in this case there are no extra fields except a doublet and singlet that can be identified with the SM fields.
We will consider all the four possible cases  with regards to the twisting and localization of the mixings, and we will discuss some interesting limits for the light mass eigenstate.
For untwisted boundary conditions ($\eta' = 1$), the master equations are:

\begin{multline} \label{Y3ds}
\mathcal{Y}_{3} (w) = \left( \cos w - \cos (2 \pi \alpha) \right)^2 + 2 \frac{\epsilon_L^2 + \epsilon_R^2}{w} \sin w \left( \cos w - \cos (2 \pi \alpha) \right) +\\
- \frac{4 \epsilon_L^2 \epsilon_R^2}{w^2}\, \cdot \left\{ \begin{array}{ll}
\left( \cos w + 1 \right) \left( \cos w - 1 + 2 \frac{w^2}{w^2+\kappa^2} \sin^2 (\pi \alpha) \right) & \mbox{different branes}\,,\\
\frac{1}{2} \left( \cos 2 w - 1 + 2 \frac{w^2}{w^2+\kappa^2} \sin^2 (2 \pi \alpha) \right)  & \mbox{same brane}\,.
\end{array} \right.
\end{multline}
where $w^2 = (2 \pi R m)^2 - \kappa^2$, and $\kappa = 2 \pi R M$.
As expected the only difference between the different branes and same brane cases is in the ``interference'' term.
For twisted boundary conditions ($\eta' = - 1$), the equation becomes:

\begin{multline} \label{tildeY3ds}
\mathcal{\tilde Y}_{3} (w) = \left( \cos w + \cos (2 \pi \alpha) \right)^2 + 2 \frac{\epsilon_L^2 + \epsilon_R^2}{w} \sin w \left( \cos w + \cos (2 \pi \alpha) \right) +\\
- \frac{4 \epsilon_L^2 \epsilon_R^2}{w^2}\,
 \cdot \left\{ \begin{array}{ll}
\left( \cos w - 1 \right) \left( \cos w + 1 - 2 \frac{\kappa^2}{w^2+\kappa^2} \sin^2 (\pi \alpha) \right) & \mbox{different branes}\,,\\
\frac{1}{2} \left( \cos 2 w - 1 + 2 \frac{w^2}{w^2+\kappa^2} \sin^2 (2 \pi \alpha) \right)  & \mbox{same brane}\,.
\end{array} \right.
\end{multline}
In the case of vanishing $\epsilon$'s, eqs~(\ref{Y3ds}) and~(\ref{tildeY3ds}) simplify to

\begin{equation}
\cos w \mp \cos (2 \pi \alpha)=0\,,
\end{equation}
where the signs refer to the untwisted/twisted case.
The solutions are

\begin{equation}
m_n^2 = M^2 + \left\{ \begin{array}{l}
\frac{(n + \alpha)^2}{R^2} \quad \mbox{untwisted}\,,\\
\frac{(n + 1/2 + \alpha)^2}{R^2} \quad \mbox{twisted}\,,
\end{array} \right. \end{equation}
as we expect.

We can solve Eqs.~(\ref{Y3ds}) and~(\ref{tildeY3ds}) approximately
 in the limit of small Higgs VEV $\alpha$: this limit is useful to understand how the effective Yukawa coupling depends on the parameters in the model.
In the four different cases we find that the ratio between the fermion mass and the $W$ mass ($\alpha$) is

\begin{eqnarray}
\mbox{untw. d.:}\; \frac{m_q}{m_W}& = &\epsilon_L \epsilon_R \frac{\kappa \coth \frac{\kappa}{2}}{\sqrt{\left( 2\epsilon_L^2 \cosh \frac{\kappa}{2} + \kappa \sinh \frac{\kappa}{2} \right) \left( 2\epsilon_R^2 \cosh \frac{\kappa}{2} + \kappa \sinh \frac{\kappa}{2} \right)}}\,, \label{eq:untwd}\\
\mbox{untw. s.:}\; \frac{m_q}{m_W}& = &\epsilon_L \epsilon_R \frac{\kappa\, \mbox{cosech}\, \frac{\kappa}{2}}{\sqrt{\left( 2\epsilon_L^2 \cosh \frac{\kappa}{2} + \kappa \sinh \frac{\kappa}{2} \right) \left(2 \epsilon_R^2 \cosh \frac{\kappa}{2} + \kappa \sinh \frac{\kappa}{2} \right)}}\,, \label{eq:untws}\\
\mbox{tw. d.:}\; \frac{m_q}{m_W}& = &\epsilon_L \epsilon_R \frac{\kappa \tanh \frac{\kappa}{2}}{\sqrt{\left(2 \epsilon_L^2 \sinh \frac{\kappa}{2} + \kappa \cosh \frac{\kappa}{2} \right) \left( 2\epsilon_R^2 \sinh \frac{\kappa}{2} + \kappa \cosh \frac{\kappa}{2} \right)}}\,, \label{eq:twd}\\
\mbox{tw. s.:}\; \frac{m_q}{m_W}& = &\epsilon_L \epsilon_R \frac{\kappa\, \sech \frac{\kappa}{2}}{\sqrt{\left( 2\epsilon_L^2 \sinh \frac{\kappa}{2} + \kappa \cosh \frac{\kappa}{2} \right) \left( 2\epsilon_R^2 \sinh \frac{\kappa}{2} + \kappa \cosh \frac{\kappa}{2} \right)}}\,. \label{eq:tws}
\end{eqnarray}

It is interesting to study such results in different limits in the bulk mass $\kappa$.
For large $\kappa$, the bulk fermion twisting becomes irrelevant: the masses of the KK modes are dominated by the bulk mass $\kappa$.
We find a different limit depending on whether the localized masses are on the same or opposite branes:

\begin{eqnarray}
\mbox{diff. branes} & \rightarrow & \frac{4 \epsilon_L \epsilon_R}{\sqrt{(2 \epsilon_L^2+1)(2 \epsilon_R^2+1)}}\; \frac{\kappa}{2}\,  e^{-\kappa/2}\,, \\
\mbox{same brane} & \rightarrow & \frac{4 \epsilon_L \epsilon_R}{\sqrt{(2 \epsilon_L^2+1)(2 \epsilon_R^2+1)}}\; \kappa\, e^{-\kappa}\,.
\end{eqnarray}
In both cases the effective Yukawa is exponentially suppressed: the two different powers are easily understood.
If the interactions are on different branes, in order to feel both the mixings, the massive fermion has to propagate from one brane to the other, thus developing a suppression of order $\exp (- \pi R M)$.
In the case of same brane localization, the fermion has to propagate to the other brane and back, therefore accumulating a double suppression.
From these limits, it is clear that we can fit the light fermions very easily, either with a large bulk mass or with small localized mixings.

The limit for small $\kappa$ is more interesting.
In the untwisted case, we find that $m_q \rightarrow m_W$.
The rationale is again simple: when the bulk mass vanishes, the two bulk fermions decouple.
The localized fields mix with one of the two bulk fermions, $\Psi$, giving a mass to the zero modes of order $\epsilon$.
The other fermion $\tilde \Psi$, on the other hand, has a light mode whose mass is exactly $\alpha/R$, as discussed in
Section \ref{sec:model}.
We can confirm these result if we expand for small $\kappa$, without any assumptions on the Higgs VEV.
The result is that for both same and different brane we find a light mode with mass $m_q R \pi \sim \sin \pi \alpha$, and a mode:

\begin{eqnarray}
\mbox{diff. branes} & \rightarrow & m_q \pi R = \frac{\epsilon_L \epsilon_R}{\sqrt{(1+\epsilon_L^2)(1+\epsilon_R^2) - \cos^2 \pi \alpha}}\,,\\
\mbox{same brane} & \rightarrow & m_q \pi R = \frac{\epsilon_L \epsilon_R \cos \pi \alpha}{\sqrt{(1+\epsilon_L^2)(1+\epsilon_R^2) - \cos^2 \pi \alpha}}\,.
\end{eqnarray}
In the twisted case the situation is more complicated: in the case of mixings on the same brane, the bulk fields that enter the mixing terms are coming from the same bulk field $\Psi$.
The other bulk field decouples, while the localized zero modes develop a mass via the massive modes in $\Psi$.
Indeed, we find only one light mode, with mass:

\begin{equation}
m_q \pi R = \frac{\epsilon_L \epsilon_R \sin \pi \alpha}{\sqrt{(1+\epsilon_L^2)(1+\epsilon_R^2)-\sin^2 \pi \alpha}} \rightarrow \frac{\epsilon_L \epsilon_R}{\sqrt{(1+\epsilon_L^2)(1+\epsilon_R^2)}}\, \alpha\,,
\end{equation}
where the limit for small $\alpha$ agrees with the limit of Eq.~(\ref{eq:tws}).
Even if we do not have a bulk mass, we can achieve again small values with small mixings.
In the case of twisted bulk fermions on different brane, expanding Eq.~(\ref{eq:twd}), we find:

\begin{equation}
\frac{m_q}{m_W} = \frac{\epsilon_L \epsilon_R}{\sqrt{(1+\epsilon_L^2)(1+\epsilon_R^2)}} \frac{\kappa}{2}\,.
\end{equation}
The mass vanishes when $\kappa \rightarrow 0$: the reason is that the localized fermions couple to different bulk fields, so they will not get a mass in the absence of a bulk mass.

\begin{figure}[tb]
\begin{center}
\includegraphics[width=10cm]{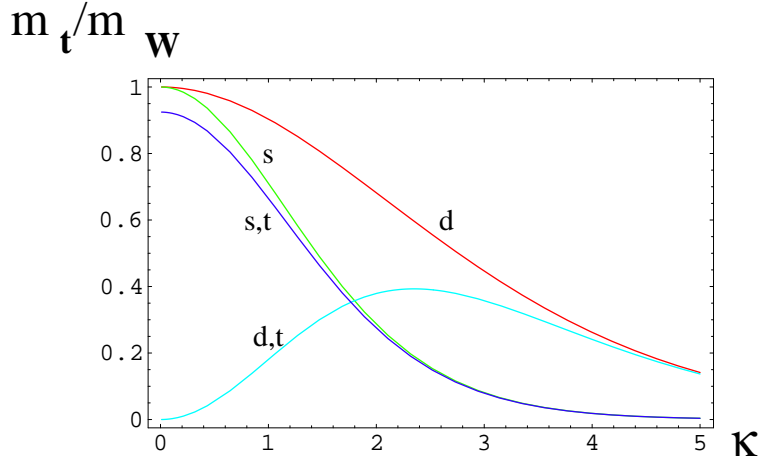}
\end{center}
\caption{Plot of the ratio $m_q/m_W$ as a function of the bulk mass
$\kappa$ with fixed mixing parameters $\epsilon_L=5=\epsilon_R$. The
four curves correspond to untwisted boundary conditions and mixings
on different branes (d) or on the same one (s), and to twisted
boundary conditions (t).} \label{fig:top}
\end{figure}

From this analysis it is clear that the effective Yukawa coupling cannot be larger than the gauge coupling.
Moreover, the only way to avoid an exponential suppression is to have a vanishing bulk mass.
In other words, we need bulk zero modes to get directly their mass via the Higgs mechanism.
If we consider larger representations, the only novelty is the presence of more states that will mix via the Higgs and complicate the equation.
However the same qualitative behavior appears.
Another interesting feature is the appearance of group theory factors that may increase the effective Yukawa coupling.
We will discuss in a later section if it is possible to use large representations to enhance the top mass.
In the rest of this section, we will focus our interest on the Higgs mass, thus the next step is to compute the one loop potential in presence of localized mixings.
Regarding the top mass, following~\cite{SSS}, for the moment we will assume it is generated by a symmetric representation.
In this case we only get a group theory factor enhancement of $\sqrt{2}$: numerically the top mass will be lower that 100 GeV.

Since
the spectrum of KK modes is largely modified in the presence of localized mixings,
we also expect large corrections in the Higgs potential.
If the mixings are small, as it may be for light fermions, we can neglect such effect and only consider the bulk contribution computed in the previous section, however a new calculation of the potential is necessary for the top quark.
Following Goldberger and Rothstein in~\cite{GR}, we can relate the Coleman-Weinberg potential to the master equation in eqs.~(\ref{Y3ds})-(\ref{tildeY3ds}).
In general, if $\mathcal{Y} (m)=0$ determines the spectrum, the contribution of the states in such a spectrum is given by:

\begin{equation}
V_{eff} = \frac{1}{2} \int_0^\infty \frac{d^4 p}{(2 \pi)^4} \ln \mathcal{Y} (i p)\,.
\end{equation}

In our case, the contribution can be written in the form of eq.~\ref{eq:potgen}, with:

\begin{equation} \label{eq:potloc}
\mathcal{F}_\epsilon (\kappa, \alpha) = \frac{1}{8} \int_\kappa^\infty d \zeta \zeta (\zeta^2-\kappa^2) \ln \frac{\mathcal{Y} (i \zeta)}{K (\zeta)}\,,
\end{equation}
where the function $\mathcal{K}$ has been added to regularize the
divergence of the integral for large $\zeta$. This function is
somewhat arbitrary, but will not affect the Higgs physics as long as
it does not depend on the Higgs itself: it will simply regularize
the divergent contribution to the vacuum energy. For instance, in
the case of eq.~\ref{Y3ds} and~\ref{tildeY3ds},

\begin{equation}
K = \cosh^2 \zeta \left( 1+\frac{2 \epsilon_L^2}{\zeta} \right) \left( 1+\frac{2 \epsilon_R^2}{\zeta} \right)\,,
\end{equation}
will ensure the exponential convergence of the integral.

\section{Higgs mass}
\label{sec:5Dmass}
\setcounter{equation}{0}
\setcounter{footnote}{0}

In this section we will finally study the dynamical determination of the Higgs VEV via the radiative potential.
The $W$ mass is given by:

\begin{equation}
m_W = \frac{\alpha}{R}\,.
\end{equation}
The value of the Higgs VEV $\alpha$ determines the ratio between the $W$ mass and the scale of the gauge boson resonances $1/R$.
Thus, the smaller $\alpha$, the heavier the resonances and the scale of new physics.

The Higgs mass is also given by the radiative potential after we expand around the VEV
\begin{equation}
\alpha = \alpha_{min} + \frac{h R}{2}\,,
\end{equation}
through the formula

\begin{equation}
m_h^2 = \frac{g_4^2}{4} R^2 V'' (\alpha_{min}) = \frac{g_4^2}{128 \pi^6} \frac{1}{R^2} \sum \mathcal{F}'' (\alpha_{min})\,.
\end{equation}
For the moment we will focus on the model in~\cite{SSS}, and only consider the bulk fermions in Table~\ref{tab:pot}.
The gauge fields alone will preserve the gauge symmetries and their potential will keep $\alpha=0$.
On the other hand, the fermion content plays a very important role, as they will generate a non trivial minimum in the potential, thus driving EWSB.
It is particularly relevant that the larger the representation the more minima we have, in particular at small values of $\alpha$.
For example, a fundamental {\bf 3} will have a minimum at $\alpha = 1/2$: this is a bad value, where the $Z$ is massless so there is an extra unwanted unbroken U(1).
The symmetric {\bf 6} will have a local minimum at $\alpha \sim 1/4$, and so on.
Another interesting point is that, twisting the boundary conditions for the fermions, we can reverse the sign of
the potential.
This feature is useful if we want to move the minimum towards small values.
The game we want to play here is to combine the contribution from different bulk fermions in Table~\ref{tab:bulkf}, the ones that are responsible for the SM fermion masses, and obtain cancellations that ensure a small value of the minimum and a heavy Higgs mass~\cite{contino2}.

In~\cite{SSS} the authors considered a well motivated scenario: the mixing terms are assumed to be all of the same order, and the hierarchy in the SM fermion masses is generated by different bulk masses.
As a consequence, the only bulk fermion giving a sizable contribution to the Higgs potential is the top.
As already mentioned, this leads to a minimum at large values of $\alpha$ and small Higgs masses.
We typically find $\alpha \sim 0.3$ and $m_h \sim 0.2 \div 0.3\, m_W$, confirming the results in~\cite{SSS}.
Another drawback of a large VEV is that it predicts a low scale for the new physics, $1/R \sim 3\div 5\, m_W = 250 \div 400$ GeV.
It is interesting to note that the only way to get a realistic value for $\alpha$ with only periodic fermions is to use a huge representation~\cite{SSS} that will certainly spoil the perturbative stability of the theory.

\begin{figure}[tb]
\begin{center}
\includegraphics[width=10cm]{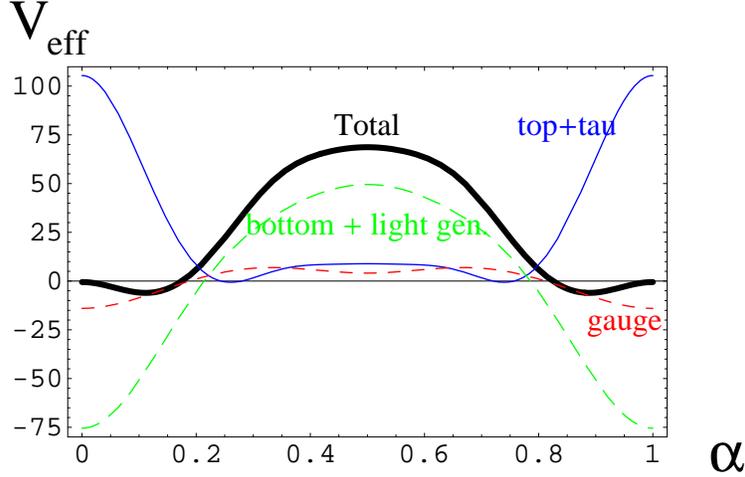}
\end{center}
\caption{Plot of the Higgs potential (in arbitrary units) from the gauge bosons (\textcolor{red}{dashed-red}), tau and top (\textcolor{blue}{blue}), twisted fermions (\textcolor{green}{dashed-green}), and the total ({\bf thick black}), for one light generation with $\kappa_l = 3$. The other parameters are like described in the text.} \label{fig:plotpot}
\end{figure}

In order to lower the value of $\alpha_{min}$ and enhance the Higgs mass,
we consider a different scenario: giving up the motivation to explain the fermion mass hierarchies with bulk masses, we assume
that light fermions are suppressed by small mixings.
In this case the bulk fermions responsible for their masses can contribute to the Higgs potential.
Moreover, we can also twist the boundary conditions for some of them, in order to achieve cancellations in the potential.
The potential will now depend on a throng of new parameters: we will concentrate on a particularly successful example.
We assume that the top {$\bf \bar 6$} has large mixing terms, say $\epsilon_L = \epsilon_R = 3$, and $\kappa_t \sim 1$.
With these numbers, we have a top around $1.3\, m_W \sim 100$ GeV.
For the bottom we add a twisted {\bf 3} with $\kappa_b = 0$.
For the tau, we include a {\bf 10} with $\kappa_\tau = 1$.
Finally, we add twisted fermions for the light generations (namely a {\bf 3}, a {$\bf \bar 6$} and a {\bf 10}), with a common bulk mass $\kappa_l$ for simplicity, that we keep as a free parameter.
In Fig.~\ref{fig:plotpot}, we plotted the single contributions to the potential for $\kappa_l = 3$.
The role played by the individual  terms is quite clear: the top and tau contributions will drive EWSB.
On the other hand, the twisted fermions, those introduced for the bottom and light generations, tend to move the minimum back to $\alpha=0$: the cancellation between these two terms allows us to get a low minimum.

\begin{figure}[tb]
\begin{center}
\includegraphics[width=6cm]{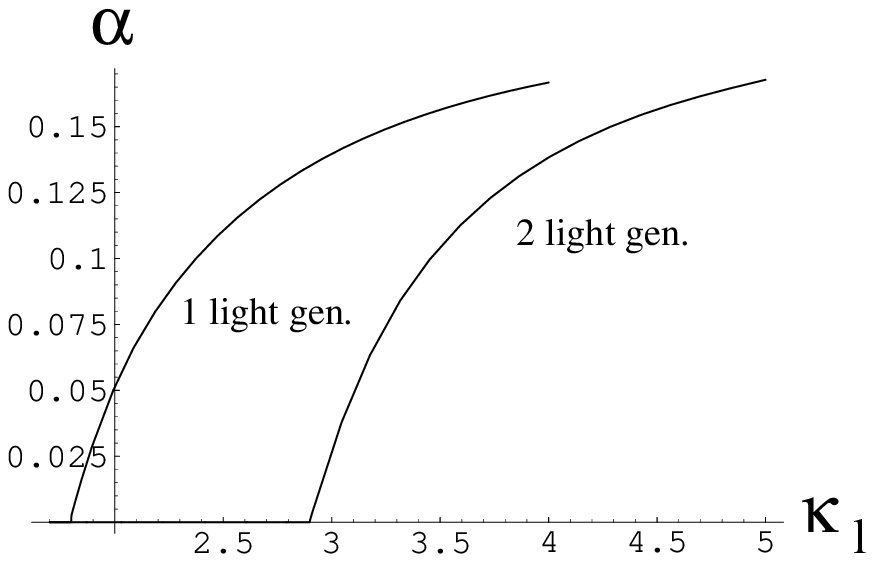} \hspace{1cm} \includegraphics[width=6cm]{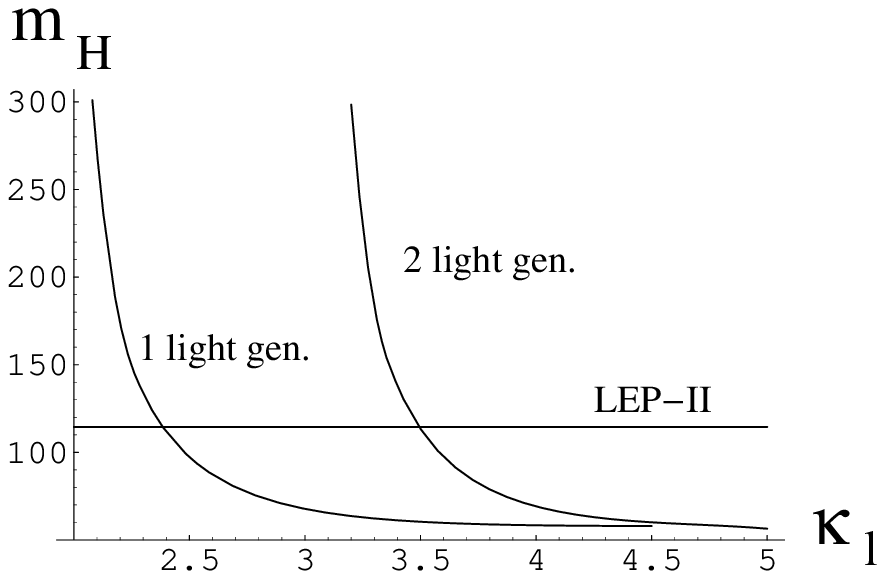}
\end{center}
\caption{Plots of the minimum  of the Higgs potential $\alpha$ (left) and the Higgs mass (right), as a function of $\kappa_l$.} \label{fig:higgsmass}
\end{figure}

In Fig.~\ref{fig:higgsmass} we show the Higgs mass and the Higgs VEV as a function of $\kappa_l$ in the case of one or two light generations taken into account.
It is important to notice that for small values of $\kappa_l$, the contribution of the twisted fermions will dominate and force the Higgs in the symmetric phase.
In other words, there is a continuous transition to the value $\alpha = 0$: however, when $\alpha$ is too small the dependence on the parameter $\kappa_l$ is very strong, signaling a fine tuning in the potential.
Values of $\alpha \gtrsim 0.1$ can be achieved without fine tuning as the cancellation only depends on the representations included in the calculation. 
We have checked that the results are also insensitive to variations of the other continuous parameters.
The LEP bound on the Higgs mass pushes $\alpha \lesssim 0.05$, in a region that shows a mild fine tuning.
We will be more quantitative in the next section, where we analyze a more realistic scenario including the top mass.
Moreover, in this region of the parameter space, the scale of the KK resonances is $1/R \gtrsim 20\, m_W \sim 2$ TeV.

\section{Top mass from large representations of SU(3)}
\label{sec:topmass}
\setcounter{equation}{0}
\setcounter{footnote}{0}

The model presented in the last sections seems to have a realistic spectrum, except for the top that is too light.
The main reason is that the Yukawa is generated by the gauge interactions themselves, so the fermion masses
have an upper bound
\begin{equation}
m_q \leq k\, m_W\,,
\end{equation}
where the bound is saturated when the bulk mass vanishes, or in
other words the fermion is a bulk field zero mode, and the
proportionality factor $k$ depends on the representation the top is
embedded in. In order to fit the top mass, one should include in the
theory a large representation. But how large does it have to be in
order to generate a realistic mass? It turns out that the number $k$
is given by the square root of the rank of the
representation~\cite{SSSflavor}, i.e. the number of indices. So, the
smallest useful representation would be a tensor with 4 indices. It
would lead to the nice prediction that at tree level $m_t = 2\, m_W$,
then corrections from QCD loops could account for the
extra enhancement. The first worry about the use of large reps is
that the cutoff of the theory may enter a strong coupling regime at low scales, due to large group theory factors
in the fermion loops. We will comment on this issue later, and for the
moment just assume that the theory is under perturbative control.

The representations of rank 4 of $SU(3)_w$, with their decomposition under SU(2)$\times$U(1), are:

\begin{equation} \begin{array}{lcl} \label{reps4}
({\bf \bar {15}})_{-2/3} &  \rightarrow & ({\bf 1}, 2/3) + ({\bf 2}, 1/6) + ({\bf 3}, - 1/3) + ({\bf 4}, -5/6 ) + ({\bf 5}, -4/3)\,, \\
({\bf \bar {24}})_{0} &  \rightarrow & ({\bf 1}, 2/3) + ({\bf 2}, 1/6) + ({\bf 2}, 7/6) + ({\bf 3}, - 1/3) + ({\bf 3}, 2/3) + ({\bf 4}, -5/6 ) \\
 & & + ({\bf 4}, 1/6 ) + ({\bf 5}, -1/3)\,, \\
({\bf 27})_{2/3} &  \rightarrow & ({\bf 1}, 2/3) + ({\bf 2}, 1/6) + ({\bf 2}, 7/6) + ({\bf 3}, - 1/3) + ({\bf 3}, 2/3) + ({\bf 3}, 5/3) \\
 & &  + ({\bf 4}, -5/6 ) + ({\bf 4}, 1/6 ) + ({\bf 5}, 2/3)\,,
\end{array} \end{equation}
where we have added a charge under the extra U(1)$_X$ in order to fix the hypercharges.
In the following we will use the smallest of these representations, the symmetric ${\bf \bar {15}}$.
In order to fully exploit the factor of two, we need the top to be a bulk fermion zero mode.
So, we only add one bulk field, and as usual
use the orbifold parity to obtain a chiral spectrum of zero modes.
The orbifold projection will leave unwanted zero modes in the large representations of SU(2)$_L$ as well: in order to get rid of them we add localized fermions, and a mass like in Eq.~(\ref{eq:lloc}).
For simplicity, we will assume in the following that all three brane localized mass
parameters are equal (and denote them by $\epsilon$).

\begin{figure}[tb]
\begin{center}
\includegraphics[width=10cm]{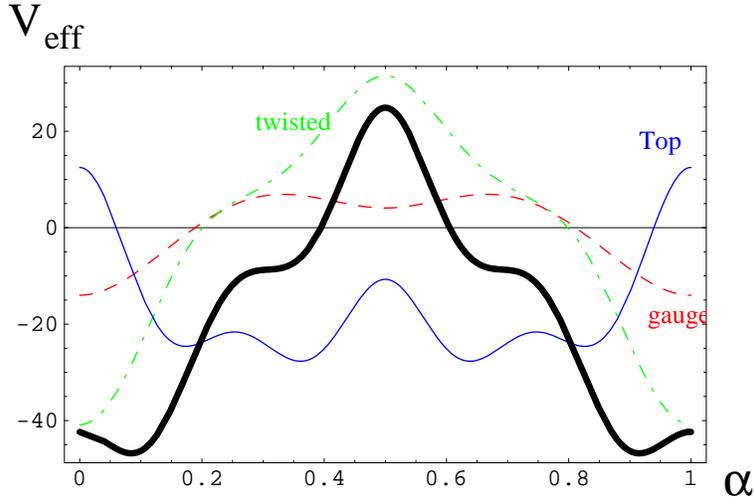}
\end{center}
\caption{Plot of the Higgs potential (in arbitrary units) from the gauge bosons (\textcolor{red}{red-dashed}), top (\textcolor{blue}{blue}), bottom ({\bf 3}) and tau ({\bf 10}) (\textcolor{green}{green-dashed}), and the total ({\bf thick black}), for $\epsilon=1.25$.} \label{fig:plotpot15}
\end{figure}

\begin{figure}[tb]
\begin{center}
\includegraphics[width=15cm]{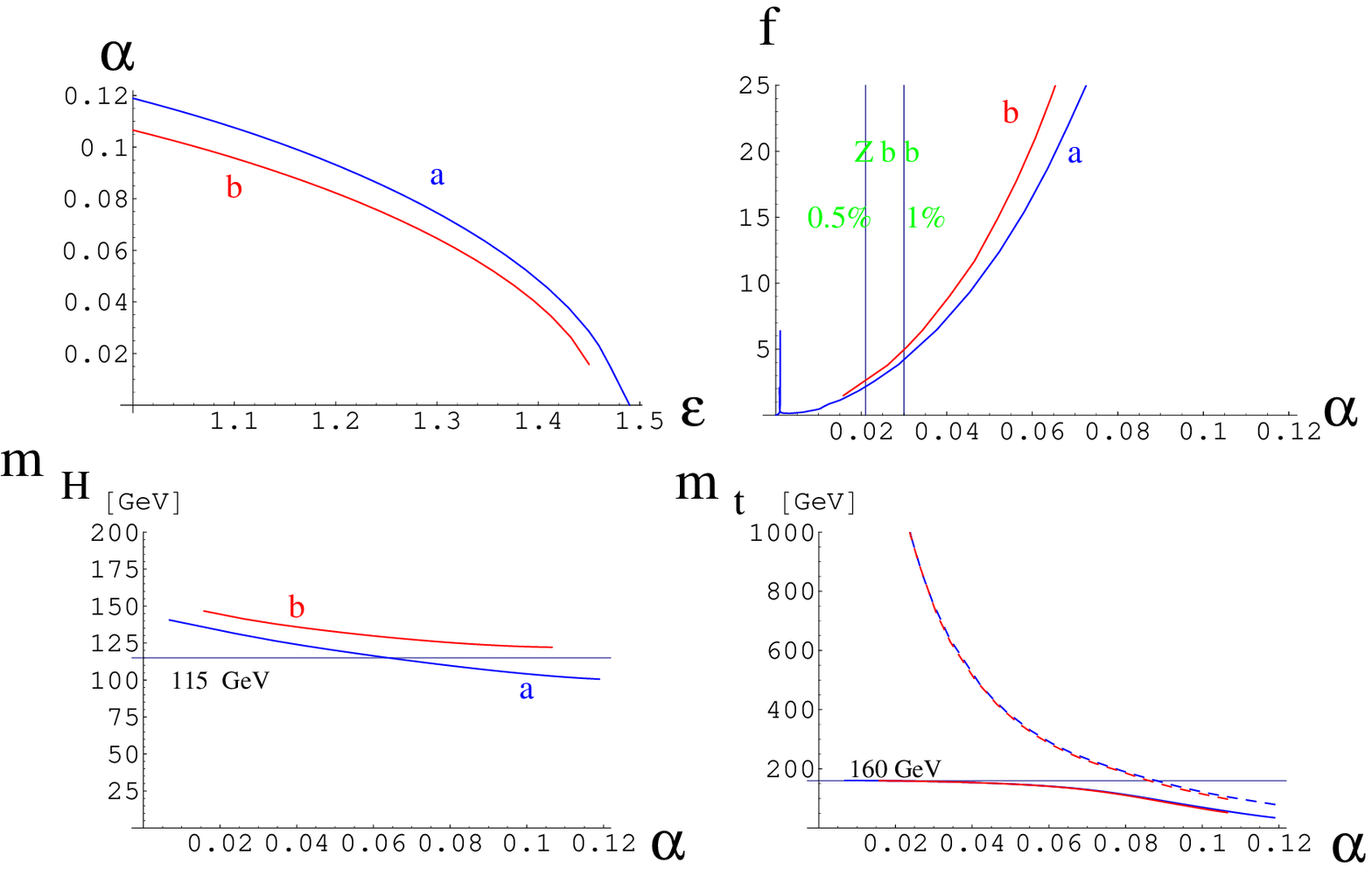}
\end{center}
\caption{Top left: Higgs VEV $\alpha$ as a function of the localized masses $\epsilon$.
Top right: fine tuning as a function of $\alpha$. The vertical lines are bounds on $\alpha$ coming from corrections to the $Z b \bar b$ vertex, discussed in Section~\ref{sec:EWPT}.
Bottom left: Higgs mass as a function of $\alpha$. The horizontal line is the LEP bound of $115$ GeV.
Bottom right: top and first massive mode (dashed) masses as function of $\alpha$. In all the plots, we show the results for model a (\textcolor{blue}{blue}) and model b (\textcolor{red}{red}).} \label{fig:top15}
\end{figure}

In order to determine if the spectrum is realistic, we need again to
minimize the Higgs potential. We analyze a minimal scenario, where
only the third generation fermions contribute. The masses for bottom
and tau are generated in the usual way, with localized degrees of freedom that feel EWSB via the mixing to massive bulk fields.
The simplest choice is to add two fermions with twisted boundary condition: they will both give them small masses and induce cancellations in the Higgs potential as in the previous section.
Due to the presence of the extra U(1)$_X$ we can always adjust the overall hypercharge, so we are free to use any representation: however, we will concentrate on small representations, like the {\bf 3}, {\bf 6} and {\bf 10} in Table~\ref{tab:bulkf}, so that they will not worsen the problem with the low strong coupling scale.
For simplicity we will present results from two cases, that summarize the qualitative properties of this scenario:

\begin{center}
\begin{tabular}{||l||c|c||}
\hline
\hline
 & bottom & tau  \\
\hline
model \textcolor{blue}{a} & ({\bf 3}, {\bf 3})$_0$ & ({\bf 1}, {\bf 10})$_0$ \\
model \textcolor{red}{b} & ({\bf 3}, {\bf 6})$_{1/3}$ &  ({\bf 1}, {\bf 3})$_{-2/3}$\\
\hline
\hline
\end{tabular} \end{center}
where we have eventually assigned U(1)$_X$ charges to fix the hypercharge of the doublet and singlet components.
A typical shape for the
Higgs potential is given in Fig.~\ref{fig:plotpot15}, for model a: EWSB is again
induced by the contribution of the top (${\bf \bar {15}}$), while the twisted
fermions push the VEV towards small values. The only free parameter
in this case is $\epsilon$: in the top--left panel of
Fig.~\ref{fig:top15} we show the value of the Higgs VEV as a function of $\epsilon$ in the two cases. 
You can see that EW symmetry is broken for natural values of $\epsilon \sim 1$.
For large
localized masses, $\epsilon \gtrsim 1.5$, there is no EWSB. What happens is
the following: the zero modes from the unwanted states are removed,
and this is equivalent to twisted boundary conditions for such
components of the bulk fermion. The spectrum of the ($\bf \bar {15}$) in this limit is
equivalent to a twisted fermion with large mixing with a localized
doublet and singlet. However, the induced potential will resemble
the one of a twisted bulk fermion as well, thus it will not drive
EWSB. For this reason we did not consider this possibility. 
In the plot, we also see that the sensitivity of the minimum to $\epsilon$ becomes large when it approaches the EW preserving values.
This signals that in that region a parametric fine tuning is at work in the potential.
A simple way to quantify the amount of fine tuning $f$ is via the logarithmic derivative:

\begin{equation}
f = \frac{d \log \alpha (\epsilon)}{d \log \epsilon}\,.
\end{equation}
which we plotted in the top--right panel of Fig.~\ref{fig:top15}.
Values of $\alpha \gtrsim 0.05$ require a fine tuning milder than 10\%: this shows that we can naturally obtain values of $1/R \lesssim 2$ TeV.

\begin{table}[tb]
\begin{center}
\begin{tabular}{||l||c|c|c|c|c||}
\hline
\hline
$\alpha$ & $1/R$ & $f$ & $m_H$ & $m_t$ & $m'_t$ \\
\hline
\hline
0.08 & 1 TeV & $\tabarr{31\%}{42\%}$ & $\tabarr{110}{125}$ GeV  & $\tabarr{113}{110}$ GeV & $\tabarr{189}{186}$ GeV \\
\hline
0.05 & 1.6 TeV & $\tabarr{11\%}{14\%}$ & $\tabarr{120}{133}$ GeV &  $\tabarr{149}{149}$ GeV & $\tabarr{381}{375}$ GeV \\
\hline
0.04 & 2 TeV  & $\tabarr{7\%}{9\%}$ & $\tabarr{124}{136}$ GeV & $\tabarr{154}{154}$ GeV & $\tabarr{519}{514}$ Gev\\
\hline
0.03 & 2.7 TeV  & $\tabarr{4\%}{5\%}$ & $\tabarr{128}{140}$ GeV & $\tabarr{157}{157}$ GeV & $\tabarr{753}{746}$ Gev\\
\hline
0.02 & 4 TeV  & $\tabarr{2\%}{2\%}$ & $\tabarr{134}{144}$ GeV & $\tabarr{159}{159}$ GeV & $\tabarr{1224}{1213}$ Gev\\
\hline
\hline
\end{tabular} \end{center}
\caption{Higgs ($m_H$), top ($m_t$) and first massive fermion ($m_t'$) masses for different values of the Higgs VEV $\alpha$, in the two models \textcolor{blue}{a}\, (top row) and \textcolor{red}{b}\, (bottom row). We also list an estimate of the fine tuning $f$ required to obtain these minima.}
\label{tab:res15}
\end{table}

In the bottom--left panel of Fig.~\ref{fig:top15}, we show the mass of the
Higgs as a function of $\alpha$ in the two cases.
In model a, the LEP bound of $115$ GeV requires that $\alpha \lesssim 0.06$: this means that in order to push the Higgs mass above the direct bound, we need to allow a moderate fine tuning in the potential.
On the other hand, in model b no such fine tuning is needed: the reason is that the colored bulk fermion is in a larger representation, so it will enhance the loop induced quartic term.
In general, $m_h \sim 120 \div 150$ GeV can be obtained, where the precise value depends on the choice of bulk fermions.
Finally, in the bottom--right panel we plotted the top mass and the mass of the first massive mode.
The results are the same in the two cases, as they only depend on the bulk $\bf \bar {15}$.
The top, in order to saturate the value $2\, m_W$, also prefers small values for $\alpha \lesssim 0.04$, where the first massive mode is heavier than $\sim 500$ GeV.
These modes are bound by direct searches of a fourth generation of quarks to be heavier than $\sim 200\div 300$ GeV (for a $b'$ the bound is $\sim 200$ GeV~\cite{PDG}, and we only expect group theory factors coming from the different representations).

In Table~\ref{tab:res15} we list some numerical results for different choices of the Higgs VEV.
In the next section, we will analyze possible constraints on the parameter space.

\section{Tree level corrections to electroweak precision observables} 
\label{sec:EWPT}
\setcounter{equation}{0}
\setcounter{footnote}{0}

Models of EWSB in extra dimensions generically obtain tree level corrections to electroweak precision observables, generated by the wave function overlap that affects the couplings between particles.
One possible source is the mixing between zero modes and KK modes of gauge bosons or fermions, generated by the Higgs VEV responsible for EWSB.
This is the case in models of Gauge-Higgs unification in warped background~\cite{contino1,Zbb}.
In a flat background, the flatness of the $W$ and $Z$ wave functions and of the Higgs profile ensures the
absence of such corrections.
The reason behind this is that the Higgs VEV does not mix the zero modes with the KK resonances, thanks to the
orthogonality of their wave functions.
The KK modes will generate non vanishing corrections at loop level, but we expect them to be small due to the heaviness of the scale of new physics.
The only tree level corrections in the flat case can be generated by the presence of exotic zero modes that mix with the SM particles and pick up a mass via boundary terms.

A source of such deviations is the presence of large SU(2)$_L$ representations among the bulk fermions needed to generate masses for the SM fermions.
Their eventual zero modes will mix with the SM fermions via the Higgs VEV.
For the light fermions such corrections are highly suppressed by the masses, so they are negligible.
The only worrisome coupling is $Z b_l \bar b_l$, because the left handed component of the b is in the same multiplet as the top, and is mixed with the large reps present in the $\bf \bar {15}$.
The leading contribution will arise at order $\alpha^2$.
At this order, the only relevant representations are the ones linked to the quark doublet by one Higgs insertion.
Gauge invariance allows only couplings with triplets, with hypercharge -1/3 or 2/3:

\begin{equation}
\mathcal{Y}_{-1/3}\, Q_L H^\dagger {\bf \bar 3}_{-1/3} +  \mathcal{Y}_{2/3}\, Q_L H {\bf \bar 3}_{2/3}\,.
\end{equation}
We can compute the corrections to the vertex in two simple limits.
If the localized mass is small, then only the zero mode contributes, as the KK modes do not couple and the mixing induced by the localized term is small.
In this limit, namely $\alpha \ll \epsilon \ll 1$, the contribution of the two triplets is:

\begin{equation} \label{eq:dzbb}
\Delta = \frac{\delta g}{g} = \frac{1}{1-\frac{2}{3} \sin^2 \theta_W} (\mathcal{Y}_{2/3}^2 - \mathcal{Y}_{-1/3}^2) \left( \frac{m_W}{m_3} \right)^2\,,
\end{equation}
where we have assumed that the two triplets have the same mass $m_3$.
In the limit of large localized mass, the BCs of the triplets are effectively twisted: we can then compute the contribution of a tower of twisted states.
The contribution is the same as (\ref{eq:dzbb}), with

\begin{equation}
\frac{1}{m_3^2} \rightarrow \frac{\pi^2}{3} R^2\,.
\end{equation}
In this limit, we will have a direct bound on $R$, or equivalently on $\alpha = m_W R$.

In the case of a {\bf 15}, discussed in the previous section, there is only a triplet with hypercharge $Y=-1/3$.
The effective Yukawa is $\mathcal{Y}_{-1/3} = \sqrt{3}$, and the correction is

\begin{equation} \label{eq:zbb3}
\Delta = - \frac{\pi^2}{1-\frac{2}{3} \sin^2 \theta_W} \alpha^2\, \approx - 11\, \alpha^2\,.
\end{equation}
The LEP experiments have constrained the deviation $\Delta$
to be less than about one percent~\cite{LEP}.
A one percent deviation would imply the bound
$\alpha < 0.03$ or $1/R > 2.7$ TeV (while a 0.5\% bound would imply
$\alpha < 0.021$ or $1/R > 3.9$ TeV). This bound would start pushing $\alpha$ in the
region where a few percent fine tuning in the Higgs potential is required. It is
interesting to note that the triplet with hypercharge $Y = 2/3$ will
give a positive contribution to $\Delta$, thus relaxing the
bound a little. This might be the case of larger representations like the
{\bf 24} or {\bf 27}, see Eq.~\ref{reps4}, that contain both the triplets. Note that this
correction is present also in the case of a small representation,
like the {\bf 6} considered in Sec.~\ref{sec:5Dmass}: the main
difference is that the effective Yukawa $\mathcal{Y}_{-1/3} =
1$ is smaller significantly reducing the correction to $Zb\bar{b}$.

Another source of deviations is the presence of the extra U(1)$_X$ needed to fit the weak mixing angle.
The orbifold projection breaks SU(3)$_w \times$U(1)$_X \to$ SU(2)$_L \times$U(1)$_w \times$U(1)$_X$, leaving an unwanted zero mode.
The two U(1)'s can be broken to the hypercharge by a localized Higgs mechanism or the presence of a (localized) anomaly.
In both cases, the net effect is the presence of a localized mass term for the combination

\begin{equation}
X_\mu = \frac{1}{\sqrt{3 g^2 + g_x^2}} \left( \sqrt{3} g A^8_\mu - g_x A^x_\mu \right)\,,
\end{equation}
where $A^8$ is the gauge boson of the U(1)$_w$, while the orthogonal combination can be identified with the hypercharge gauge boson

\begin{equation}
B_\mu =  \frac{1}{\sqrt{3 g^2 + g_x^2}} \left( g_x A^8_\mu + \sqrt{3} g A^x_\mu \right)\,,
\end{equation}
with gauge coupling

\begin{equation}
g' = \frac{\sqrt{3} g_x g}{\sqrt{3 g^2 + g_x^2}}\,.
\end{equation}
At leading order, $g_x$ can be tuned to fit the SM coupling of the hypercharge.
However, the presence of the $X$ boson will generate corrections to EWP observables.
The localized mass term is not protected by any symmetry, so it will generically be as large as the cutoff of the theory: in this limit it will distort the wave functions of the KK modes and introduce mixings of such massive states with the zero modes.
If the light fermions are localized on one fixed point (this is the same point where the anomaly is localized) they will not couple to the $X$ boson, thus there will not be any correction to the couplings at order $\alpha^2$: the corrections will be of order $\alpha^4$, or induced by the mixing with the bulk fermions that generate the mass, thus being of order $(m_f R)^2 \alpha^2$.
In both cases, they are safely small: this implies that the $S$ parameter is negligibly small.
However, the $X$ boson will also mix with the $Z$, correcting its mass and generating a deviation in the $\rho$ parameter (or alternatively $T$) given by

\begin{equation}
T =  \frac{4 \pi}{e^2} \Delta \rho = \frac{4 \pi}{e^2} \frac{\pi^2}{3} \frac{3 - 4 \sin^2 \theta_W}{\cos^2 \theta_W}\, \alpha^2\, \approx 1.2\cdot 10^3\, \alpha^2.
\end{equation}
The experimental bound $|T| \lesssim 0.3$ poses a bound $\alpha < 0.015$ ($1/R > 5$ TeV).

Another correction arises in the third generation sector: the $b_l$ is a bulk field, so it couples directly to the $X$ bosons.
This coupling induces a correction to the coupling with the $Z$ given by

\begin{equation}
\Delta = - \frac{2}{3} \left[ \frac{3 - 4 \sin^2 \theta_W}{\cos^2 \theta_W} \frac{1}{6} - 3 Q_x \right]\, \frac{\pi^2}{1-\frac{2}{3} \sin^2 \theta_W} \alpha^2\,,
\end{equation} 
where the U(1)$_X$ charge of the $\bf \bar {15}$ is $Q_x = -2/3$.
This contribution is negative, and when added to the correction from the triplet in Eq.~(\ref{eq:zbb3}), it gives a bound $\alpha < 0.018$ or $1/R > 4.5$ TeV for $|\Delta| < 1\%$ ($\alpha < 0.013$ or $1/R > 6$ TeV for $|\Delta| < 0.5\%$).

A potentially tight bound on the scale $1/R$ comes from the couplings of gauge resonances of SU(2)$_L \times$U(1)$_w$ with the light fermions, that are localized on the orbifold fixed points.
Their coupling is generically $\sqrt{2}\, g$, and they will induce four fermion operators at tree level.
The bound from precision electroweak observables~\cite{KKgb,LEP2} would require $1/R > 4$ TeV.
An even tighter bound, around $7$ TeV, would emerge from the analysis of LEP2 data off the $Z$ peak~\cite{cheung}.
However, the light fermions do not play any active role in the extra dimensional construction.
Moreover, their effect on the little hierarchy problem is negligible, due to the smallness of the Yukawa couplings.
We can thus couple them to the zero modes of the gauge bosons only, and add an explicit Yukawa with the Higgs without spoiling the good features of the model.
A consistent inclusion of them is postponed to a UV completion of this model.
The only fermions that play an active role in the model under discussion are the top and bottom, and the constraints in this sector are milder.
It is anyway very easy to think of a scenario where four fermion operators involving top and bottom are absent or suppressed.
For instance, we can have a bulk $\bf \bar {15}$, that contains the doublet $Q_L$ and right handed top $t_R$. After EWSB their wave functions remain flat and the top gets its mass.
The right handed bottom $b_R$ can also be identified with the zero mode of a bulk singlet of SU(3)$_w$.
In order to give a mass to the bottom, we can introduce a twisted bulk fermion (like the {\bf 6} in model b of 
Section~\ref{sec:topmass}), and mix it with the bulk $Q_L$ and $b_R$ on the branes.
For the bottom mass, it is enough to have mixings $\epsilon \sim 0.1$.
The flatness of the wave function of the bulk fields will forbid couplings of two zero modes with one KK mode, thus removing the four fermi operators at tree level.
This mechanism is precisely the extra dimensional version of the $T$ parity advocated for Little Higgs models.
The presence of localized mixings will distort the wave functions, and generate very small contributions suppressed by $\epsilon^4 \sim (m_b/m_W)^2 \sim 10^{-3}$. 
Note that the tau (the bulk {\bf 3}) does not play an important role in model b, so it can be treated as a light fermion.

Another effect of the large representations that we need to fit the top mass is that they tend to lower the scale where the theory is strongly coupled, due to their large contributions to loops.
Naive dimensional analysis would suggest that the strong coupling scale is:

\begin{equation}
\Lambda_{strong} R \sim \frac{24 \pi^3 R}{g_5^2} = \frac{24 \pi^2}{g^2} \sim 10^3\,,
\end{equation}
however this estimation does not take into account large group
theory factors. For example, a fermionic representation will
contribute to a one loop result with a factor $4 C(r)$, $C(r)$ being
the Dynkin index of the representation. 
Notice that this same factor will enhance the contribution to the Higgs mass term~\cite{matchev}.
For the {\bf 15}, the
smallest representation needed to generate a top mass, this would
mean $4 \cdot 3 \cdot 35/2 \sim 200$, where we also took into
account a color factor of $3$. Thus the naive strong scale is only
few times the resonance scale $1/R$: it is then very important to
verify if the results are stable under further radiative
corrections. For example, a calculation of the two loop corrections
to the Higgs mass would be useful. In the case of a {\bf 24}, {\bf
27} or higher rank representations, the Dynkin index is much larger,
thus spoiling the predictive power of the theory.

\section{Conclusions and Outlook}
\label{sec:concl}
\setcounter{equation}{0}
\setcounter{footnote}{0}

We have shown that in a minimal model of Gauge-Higgs unification in a flat extra dimension,
it is possible to accommodate a large Higgs mass and a heavy top. We
do not add extra fields just for the sake of the Higgs potential,
but all the bulk fields are also used to generate masses for the SM
fields: in this sense we preserve minimality. 
Tree level corrections to electroweak precision measurements generated by the mixing between KK modes are avoided thanks to the flatness of the Higgs profile.
The most serious bound on the size of the extra dimension comes from the presence of zero modes that are not projected away by the orbifold: extra zero modes in the large representation needed to fit the top mass correct the $Z b \bar b$ vertex, while the extra U(1)$_X$ needed to fit the weak mixing angle introduces a $\Delta \rho$ (or $T$).
The top representation will also lower the scale where the theory becomes strongly coupled to few $\times\,1/R$.

The model we analyzed
consists of a SU(3)$_c \times$SU(3)$_w \times$U(1)$_X$ gauge
group, where the extra gauged U(1)$_X$ fixes the value of $\sin^2
\theta_W$. The orbifold projection breaks SU(3)$_w \to$ SU(2)$_L
\times$U(1)$_w$, and U(1)$_X \times$U(1)$_w$ can be broken to
U(1)$_Y$ either by twisted boundary conditions or by a localized anomaly. The
top mass is generated via gauge interaction, so one is forced to
introduce a $\bf \bar {15}$ of SU(3)$_w$ in order to
generate an enhancement of the fermion mass with respect to the $W$
mass.
Regarding the Higgs potential, introducing twisted bulk fermions, that also
give mass to the bottom and tau, one can lower the value of the
Higgs VEV and enhance the Higgs mass itself above the experimental
bounds via cancellations in the potential.
Part of these cancellations are due to the presence of twisted fermions: they allow to achieve naturally, without fine tuning, values $\alpha \gtrsim 0.05$ and $m_h \sim 120 \div 140$ GeV.
In this region we estimate a fine tuning milder than 10 \%.
The scale of new physics is also large, naturally around few TeV.
The only tree level corrections to the electroweak precision measurements are corrections to the $Z b \bar b$ coupling, arising via the mixing of the left-handed bottom with triplets of SU(2)$_L$ contained in the top bulk $\bf \bar {15}$, and $\Delta \rho$ induced by the extra U(1)$_X$.
They push the Higgs VEV in a region of the parameter space where a few \% fine tuning is needed, and require $1/R \gtrsim 5 \div 6$ TeV.
This bound is removed if the breaking of the gauge symmetry only comes from the orbifold projection, as it may be possible if we consider different gauge groups and more complicated orbifold projections.
The deviation in $Z b \bar b$ given by the large top representation alone would require $1/R \gtrsim 4$ TeV (at $0.5 \%$).

Another consequence of the large top representation is that the theory becomes strongly coupled at a relatively small scale, of order few $\times\,1/R$.
This is a borderline situation, and a calculation or estimation of the two loop effects is needed to decide if the calculation of the Higgs mass that we presented in this paper is reliable.
This scale is nevertheless large enough, so that higher order operators generated by the non-perturbative physics will give a negligible contribution to the precision measurements.
Above the strong coupling scale, the theory is no longer under perturbative control.
If the gauge symmetry, responsible for the protection of the Higgs mass, is not broken by non-perturbative effects, the Higgs mass is protected up to the Planck scale, thus addressing the Big Hierarchy problem.
The only quantity that may be sensitive to the UV physics, and reintroduce a UV sensitivity in the Higgs mass is the compactification scale $1/R$.
However, we can imagine to add a stabilization mechanism~\cite{stabilize}, that only couples to the gauge sector via gravitational interactions, thus without spoiling the stability of the Higgs mass.
Moreover, this stabilization mechanism will induce small distortions of the background and small 5D Lorentz violating effects in the SM sector.

\section*{Note added}

While this work was completed, we became aware of a related work by G.~Panico, M.~Serone and A.~Wulzer~\cite{PSW}, where the authors also address the top and Higgs mass problem in this context.
The main difference between the two models is that they enhance the top mass via large explicit violation of 5 dimensional Lorentz invariance in the bulk, without the introduction of large representations.

\section*{Acknowledgments}
We thank Roberto Contino, Marco Serone and Andrea Wulzer for useful discussions and comments.
We thank Giuliano Panico, Marco Serone and Andrea Wulzer also for sharing with us a draft of their forthcoming paper~\cite{PSW}, and for useful comments on our manuscript.
G.C. and C.C. thank the Aspen Center for Physics for its hospitality while part of this work was completed.
The research of G.C. and C.C.
is supported in part by the DOE OJI grant DE-FG02-01ER41206 and in part
by the NSF grants PHY-0139738  and PHY-0098631. S.C.P. was supported by the Korea Research Foundation Grant funded by the
Korean Government (MOEHRD), grant KRF-2005-214-C00147.

\end{document}